\documentclass[oneside]{article}

\usepackage{arxiv}

\usepackage{times,amsmath,epsfig}
\usepackage{fancyhdr}
\usepackage{amsmath}
\usepackage{amsfonts}
\usepackage{amssymb}
\usepackage{array}
\usepackage{graphicx}
\usepackage{url}
\usepackage{subfigure}
\usepackage{bm}
\usepackage{breqn}
\usepackage{xcolor}
\usepackage{soul}
\usepackage{amssymb}
\usepackage[breaklinks=true]{hyperref}
\usepackage{breakcites}

\usepackage{booktabs}
\usepackage{multirow}
\usepackage{nicefrac}
\usepackage{algorithmic}
\usepackage[toc,page]{appendix}
\usepackage{siunitx}

\newcommand{\TP}{\text{TP}}
\newcommand{\TN}{\text{TN}}
\newcommand{\FP}{\text{FP}}
\newcommand{\FN}{\text{FN}}

\newcommand{\SE}{\text{SE}}
\newcommand{\SP}{\text{SP}}

\title{Ensemble learning using individual neonatal data for seizure detection}

\author{Ana~Borovac\\
        Faculty of Industrial Engineering, Mechanical Engineering and Computer Science\\
        University of Iceland\\
        Kvikna Medical ehf. \\
        Reykjavik, Iceland \\
        \texttt{anb48@hi.is} \And 
        Steinn~Gudmundsson \\
        Faculty of Industrial Engineering, Mechanical Engineering and Computer Science\\
        University of Iceland\\
        Reykjavik, Iceland \\
        \texttt{steinng@hi.is} \And 
        Gardar~Thorvardsson \\
        Kvikna Medical ehf.\\
        Reykjavik, Iceland \\
        \texttt{gardar@kvikna.com} \And 
        Saeed~M.~Moghadam \\
        BABA Center \\
        Helsinki University Hospital \\
        University of Helsinki \\
        Helsinki, Finland \\
        \texttt{saeed.montazeri@helsinki.fi} \And 
		Päivi~Nevalainen \\
        BABA Center \\
        Epilepsia Helsinki and Department of Clinical Neurophysiology \\
        Helsinki University Hospital \\
        University of Helsinki \\
        Helsinki, Finland \\
        \texttt{paivi.nevalainen@hus.fi} \And  
		Nathan~Stevenson \\
		Brain Modelling Group \\
		QIMR Berghofer Medical Research Institute \\
		Brisbane, Australia \\
		\texttt{nathan.stevenson@QIMRBerghofer.edu.au} \And 
		Sampsa~Vanhatalo \\
        BABA Center \\
        Helsinki University Hospital \\
        University of Helsinki \\
        Helsinki, Finland \\
        \texttt{sampsa.vanhatalo@helsinki.fi} \And 
		Thomas~P.~Runarsson \\
		Faculty of Industrial Engineering, Mechanical Engineering and Computer Science\\
        University of Iceland\\
        Reykjavik, Iceland \\
        \texttt{tpr@hi.is}
}

\begin{document}
\maketitle

\begin{abstract}
Sharing medical data between institutions is difficult in practice due to data protection laws and official procedures within institutions. Therefore, most existing algorithms are trained on relatively small electroencephalogram (EEG) data sets which is likely to be detrimental to prediction accuracy. In this work, we simulate a case when the data can not be shared by splitting the publicly available data set into disjoint sets representing data in individual institutions. We propose to train a (local) detector in each institution and aggregate their individual predictions into one final prediction. Four aggregation schemes are compared, namely, the majority vote, the mean, the weighted mean and the Dawid-Skene method. The method was validated on an independent data set using only a subset of EEG channels. The ensemble reaches accuracy comparable to a single detector trained on all the data when sufficient amount of data is available in each institution. The weighted mean aggregation scheme showed best performance, it was only marginally outperformed by the Dawid--Skene method when local detectors approach performance of a single detector trained on all available data.
\end{abstract}

\section{Introduction}

Seizures are common during perinatal period~\cite{glass2014risk}, and management of neonatal seizures requires timely detection and treatment to reduce ensuing brain damage~\cite{bjorkman2010seizures}. The current gold standard for neonatal seizure detection is visual analysis by a human expert using a full-montage video electroencephalogram (EEG)~\cite{pressler2021ilae}. Since such service is rarely available in neonatal intensive care units (NICUs), there is an urgent clinical need for automated neonatal seizure detection algorithm (NSDA) with human expert level accuracy. 

Early automated NSDAs were based on \emph{features}, quantitative descriptors of short, e.g.\ $10 - 16$~sec long, EEG segments and expert-defined threshold decision rules~\cite{liu1992detection, nagasubramanian1997line, navakatikyan2006seizure}. Hard-coded thresholds were later replaced by statistical techniques, such as linear discriminant analysis~\cite{greene2008comparison}, support vector machines (SVMs)~\cite{ahmed2017exploring, ansari2016improved, temko2011eeg} and neural networks~\cite{hassanpour2004time}. Recently, promising results have been obtained using convolutional neural networks (CNNs)~\cite{ansari2019neonatal, isaev2020attention, o2020neonatal}. 

Deep neural networks (DNNs) generally require a large amount of training data~\cite{lecun2015deep}. However, building a large and diverse enough neonatal EEG data set with high quality seizure annotations is time consuming, ambiguous~\cite{malone2009interobserver, stevenson2015interobserver} and often limited due to strict regulations (e.g.\ the Privacy Rule of the U.S.\ Health Insurance Portability and Accountability Act (HIPAA), or the European General Data Protection Regulation (GDPR)) making data sharing between institutions difficult, if not impossible~\cite{eicher2020comprehensive,yang2019federated}. Challenges in sharing data have triggered growing interest in distributed approaches to statistical learning~\cite{kirienko2021distributed}. 
 
One approach that requires minimal sharing of information is model ensembling, i.e.\ models are trained locally at each institution and predictions on new data are aggregated (ensembled) from predictions made by the local models. This requires sharing only the models across the network of institutions rather than sharing the potentially sensitive, original biosignals. However, the procedures in model sharing need to be planned so that they mitigate the impact of possible inadvertent leaks of training data through a model~\cite{fredrikson2015model, zhang2020secret}. One solution to this problem is to have a \emph{trusted agent} in charge of the models and an aggregation procedure. Compared to the federated learning~\cite{li2020federated}, ensembling does not require communication between the institutions during the training phase (which may be difficult to set up) and it does not require the institutions to use the same model architecture. One institution could e.g.\ use a DNN, another an SVM and a third a decision tree classifier.
 
Once predictions on new data have been made there are a number of techniques by which they can be ensembled. If predictions are accompanied by probabilities they can be averaged~\cite{chang2018distributed, tuladhar2020building}, if not, a commonly used method for label aggregation is to simply select the most frequent label, referred to as \emph{majority vote} in the following. One could also put more weight on some predictions if they are a priori more trustworthy, otherwise, an estimate of each annotator performance can be used~\cite{tao2018domain, tian2015max, wolpert1992stacked}. In 1979 Dawid and Skene~\cite{dawid1979maximum} used an expected maximization~(EM) algorithm~\cite{dempster1977maximum} to estimate annotator performance and provide consensus labels. 

Ensemble learning has previously been used in neonatal seizure detection. In~\cite{pan2019aeeg} stacking is used where different model types trained on the same data are combined. In~\cite{tanveer2021convolutional} three identical NSDAs are trained on the same EEG data but using labels from different experts. In this work we use ensemble learning on disjoint data sets, to simulate the situation were institutions train NSDAs on locally available data. Depending on the training data available at each institution and its similarity to new data to be labelled, the local NSDAs are expected to vary in performance. The main contribution and novelty of this work is in the discovery of how such locally trained models can be aggregated with the aim of achieving performance comparable to a single state-of-the-art NSDA trained on the union of all local training data sets. For aggregation we compared the majority vote, the mean, the weighted mean (via stacking) and the Dawid--Skene expected maximization algorithm. We show that the weighted mean outperforms the other methods if the NSDAs in the ensemble are trained on very few patients and Dawid-Skene marginally outperforms the other methods when the local NSDAs are not much worse than the state-of-the-art NSDA. The NSDAs and ensembles are further validated on an independent data set consisting of more than 2100 hours of EEG recorded from a small subset of the channels used to train the classifiers.

\section{Methods}

\begin{figure}[t]
    \centering
    \includegraphics[width = 0.7\columnwidth]{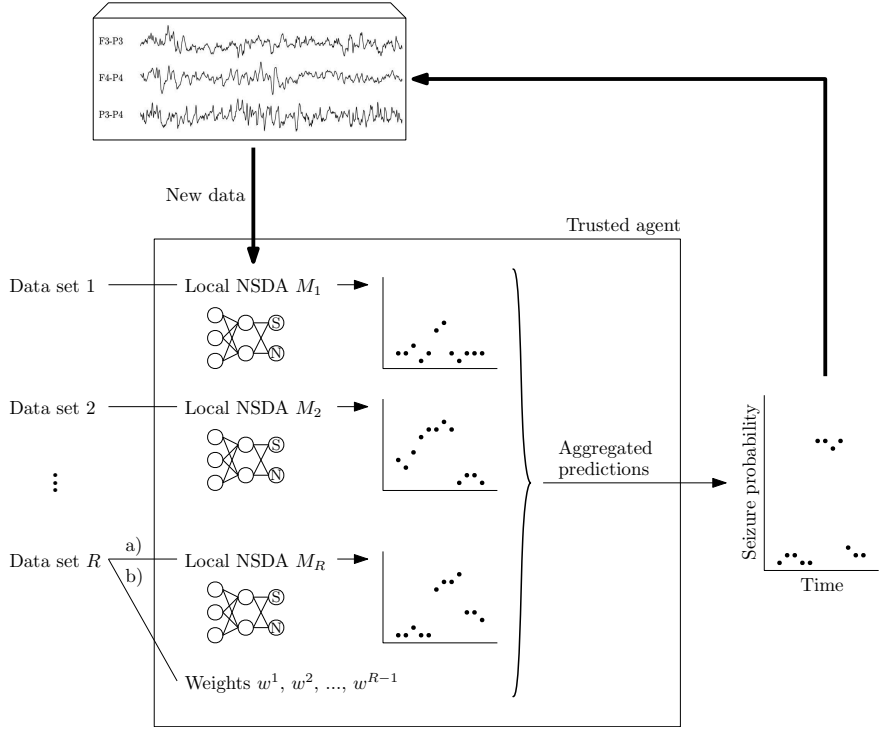}
    \caption{A schematic diagram of the proposed method. Each data set is used to train a local NSDAs or weights that are shared with a trusted agent. The trusted agent makes predictions on new data. Seizure predictions for new data are obtained a) by aggregating predictions made by $R$ NSDAs using the majority vote, the mean or the Dawid-Skene method, or, b) by aggregating predictions made by $R-1$ local NSDAs using the weighted mean (weights are learned on the $R^{\text{th}}$ data set).}
    \label{fig: schemeEM}
\end{figure}

Multiple local models, referred to as \emph{local NSDAs} in the following, are trained on disjoint subsets of multi-channel EEG recordings, simulating a scenario where several hospitals train NSDAs individually, without sharing patient data. The trained detectors are then shared with a trusted agent. To classify a short EEG segment from a new patient as seizure/non-seizure, the trusted agent sends the segment through all the local NSDAs and the predictions are aggregated using one of the following schemes: majority vote, mean, weighted mean or the Dawid--Skene method. The methodology is summarized in figure~\ref{fig: schemeEM}.

For local NSDAs, we used DNNs which take EEG segments as input. The networks share the same architecture but have different network weights since they were trained on disjoint training sets. 

\subsection{Aggregation schemes}
In the following we consider a binary classification problem where the classes are labeled 0 and 1. Let $D$ be a set of $N$ predictions from $R$ independent models
\[ D = \left\{ \left( p_1^1, p_1^2, \dots, p_1^R \right), \ldots, \left( p_N^1, p_N^2, \dots, p_N^R \right) \right\},
\]
where $p_i^j$ is the estimated probability of model $j$ of instance $i$ belonging to class 1. By setting a threshold between the classes to $0.5$, the predicted label of model $j$ of instance $i$ is given by
\[
y_i^j = 
\begin{cases}
1; & \textrm{if}~p_i^j \geq 0.5 \\
0; & \textrm{otherwise}
\end{cases}.
\]

A simple way to aggregate multiple predictions for instance~$i$, when models do not output their confidence (e.g.\ class probabilities), is to use majority vote, i.e.\ select the most frequent label. Here we use the mean of predicted labels,
\begin{equation}
    \mu_i^{M\!V} = 
    \frac{1}{R} \sum_{j = 1}^R y_i^j; \quad i \in \{1, 2, \dots, N\}.
    \label{eq: majorityVote}
\end{equation}

When the models output class probabilities, which is e.g.\ the case when the models correspond to the neural networks, the predictions can be aggregated by taking the mean probability \begin{equation}
\mu_i^M = \frac{1}{R} \sum_{j = 1}^R p_i^j; \quad i \in \{1, 2, \dots, N\}.
\label{eq: mean}
\end{equation}

As some of the models might perform better than others, a weighted mean can be used to emphasize the more accurate models. To get the final prediction in a range between 0 and 1, we used logistic regression 
\begin{equation}
\mu_i^{W\!M} = \sigma \left( \sum_{j = 1}^R w^j p_i^j \right); \quad i \in \{1, 2, \dots, N\},
\label{eq: weightedMean}
\end{equation}
where $\sigma(x) = \nicefrac{1}{(1 + e^{-x})}$. The weights for $w^j$ are learned on a held out data set (see section~\ref{subsec: training}).

The fourth aggregation method evaluated here is the Dawid--Skene method. The method estimates the sensitivity and specificity of each model, together with consensus predictions~$\mu^{D\!S}$. For details of the method see appendix~\ref{app: dawidSkene}. To predict the absence/presence of seizures from the above aggregation schemes, a threshold of 0.5 is used.

\subsection{Data}

The EEG data used to train the NSDAs is a publicly available data set containing 79 approximately one hour long neonatal EEG recordings, measured with 19 Ag/AgCl electrodes positioned according to the 10-20 system~\cite{stevenson2019dataset}. An 18 channel montage is used, i.e.\ we derive channels Fp2-F4, F4-C4, C4-P4, P4-O2, Fp1-F3, F3-C3, C3-P3, P3-O1, Fp2-F8, F8-T4, T4-T6, T6-O2, Fp1-F7, F7-T3, T3-T5, T5-O1, Fz-Cz and Cz-Pz. The recordings are annotated by three EEG experts where each second in a recording is annotated as a seizure or non-seizure. We refer to this data set as 18-channel~DS below.

The second, proprietary, data set (the 3-channel DS) consisting of EEG recordings of 28 neonates, is used as a held out test set to evaluate the aggregation schemes in a real world setting, i.e.\ detectors are trained on the 18-channel DS and tested on this data set. The data set is also used in~\cite{tapani2022validating} and is a subset of the data set used in~\cite{nevalainen2019bedside}. Institutional Research Review Board of the HUS diagnostic center approved the use of this data, including a waiver of consent due to the study’s retrospective and observational nature. Each recording spans from 19~hours to 7~days. The recordings were obtained using 4 needle electrodes (F3, F4, P3 and P4) with a common reference, instead of the full set of 19 electrodes used in the training data set. Neonatal recordings are typically performed with this reduced electrode set to allow easier maintenance in a long duration brain monitoring~\cite{boylan2013monitoring}. The three bipolar derivations (F3-P3, F4-P4 and P3-P4) are used for both two human expert annotators and as the detectors input. 

Additional attributes of the data sets are given in table~\ref{tab: basicInfo} in appendix~\ref{app: data}.

Each EEG recording is cut into 16~sec long segments with 12~sec overlap. Out of the 79 (28) recordings in 18-channel~DS (the 3-channel DS), 38 (24) contain at least one seizure longer than 16~sec identified by three (two) human experts, meaning each of these recordings contain at least one consensus seizure segment. Segments containing more than 1~sec of zero voltage interval in at least one channel (disconnected electrode or pause in the recording) are left-out from the training and test sets. The signals are filtered with a 6th order Chebyshev Type 2 band-pass filter with cut-off frequencies of 0.5~Hz and 16~Hz, down-sampled to 32~Hz and rescaled to 16-bit integers. This is similar to the pre-processing in~\cite{temko2011eeg, isaev2020attention}.

\subsection{Neonatal seizure detection algorithm}

Each NSDA is a neural network consisting of three components; a feature extractor, an attention layer and an output layer. The feature extractor is a CNN from~\cite{o2018investigating}. The features are extracted from each EEG channel separately and are combined into a single feature channel by the attention layer~\cite{isaev2020attention}. The attention layer is used since expert labels are not specific to individual channels and neonatal seizures tend to be partial~\cite{pressler2021ilae}, i.e.\ localized in a small area of the brain and therefore only present in a subset of the recorded channels. The attention layer is also independent of the number of input feature channels making the detector independent of the number of recorded EEG channels. The output layer is a fully connected layer with two output nodes representing the two classes. A detailed description of the network architecture is given in appendix~\ref{app: dnn}.

To compare the aggregation schemes to current state-of-the-art NSDAs, we trained a neural network using all the recordings in the 18-channel DS containing at least one consensus seizure longer than 16~sec ($P$). This NSDA is referred to as the \emph{baseline NSDA} in the following.

The local NSDAs use the same neural network architecture as the baseline NSDA but differ in the data used for training. The patients in $P$ (patients containing a consensus seizure) are partitioned into $k=3,4, \ldots,10$ subsets representing data sets in individual institutions. Partitioning is random such that each patient is in exactly one subset and there are at least three patients in every subset. The union of the $k$ subsets is then $P$, the data set used as a training set for the baseline NSDA. By excluding patients without consensus seizures we ensure each subset has patients with seizures and eliminated the varying number of EEGs with normal brain activity in individual subsets, making the analysis more straightforward. As there can be a big difference between the training set sizes, we obtain local NSDAs with different generalisation strengths and consequently with different performance strengths on unseen data. This is expected in practice. Even though the acquisition equipment is subject to international standards and the electrodes are positioned according to the 10-20 system, the EEG signals may vary considerably depending on the patient cohorts as the signals differ between neonates of different ages and conditions~\cite{hrachovy2015atlas, husain2005review}. Therefore, the detectors are expected to perform differently on unseen data.

\subsection{Training}    
\label{subsec: training}

After partitioning the training set, each NSDA (baseline NSDA and local NSDAs) is trained on 16~sec long EEG segments corresponding to the consensus seizures and non-seizure segments. To avoid complications due to class imbalance~\cite{isaev2020attention, johnson2019survey}, the training sets are balanced prior to training by sub-sampling the non-seizure segments. Segments with disagreements between the human experts and partly seizure/non-seizure segments are not included in the training sets. Cross entropy is used as the loss function. The Adam optimizer is used to optimize the network weights using an initial learning rate of 0.001 which is then halved every 10 epochs. The NSDAs are trained for 30 epochs with a mini-batch size of 32. Hyper-parameters, learning rate and number of epochs, are tuned empirically, from observing the behavior of the loss function during the training of the baseline NSDA. A small mini-batch size is chosen due to a small amount of data used in some local NSDAs. 
For the weighed mean aggregation scheme, the weights~$w^j$, $j \in \{1, 2, \dots, R\}$, are learned using a stacking classifier~\cite{wolpert1992stacked}. A logistic regression classifier is trained using the data from one randomly selected local NSDA in each experiment. This local NSDA is not used in an ensemble for making predictions on a test patient. Therefore, non-overlapping data sets are used for training the local NSDAs and the logistic regression classifier. Also, the training data of the local NSDAs would not need to be shared in practice as the input of the logistic regression classifier is just a set of seizure probabilities estimated by the local NSDAs and these can be provided by the trusted agent.

All the deep learning code used in the experiments is implemented using PyTorch 1.7.1~\cite{NEURIPS2019_9015} and run on an NVIDIA GTX 1080 Ti GPU. For logistic regression, we use the scikit-learn~\cite{pedregosa2011scikit} implementation with default hyper-parameters. 
The code is available at \url{github.com/anaborovac/Distributed-NSDA} (currently inactive link).

\subsection{Performance}

To avoid overlap between training and test data when evaluating classifier performance on the 3-channel DS, leave-one-subject-out cross-validation is used. This entailed training 38 baseline NSDAs, 38 sets of local NSDAs and 38 sets of logistic regression classifiers, leaving out data from one subject (patient) at a time. The experiment is repeated 10 times, resulting in $10 \cdot 38 \cdot (3 + 4 + \cdots + 10) = 19760$ local NSDAs and $10 \cdot 38 \cdot (1 + 1 + \cdots + 1) = 10 \cdot 38 \cdot 8 = 3040$ logistic regression classifiers. 

Data from each left-out patient is sent through the corresponding baseline NSDA and local NSDAs. Predictions from the baseline NSDAs are compared to human expert labels to obtain performance metrics. Predictions from the local NSDAs are first aggregated using one of the aforementioned aggregation schemes: majority vote~\eqref{eq: majorityVote}, mean~\eqref{eq: mean}, weighted mean~\eqref{eq: weightedMean} and the Dawid--Skene method (appendix~\ref{app: dawidSkene}) to obtain the final predictions and these are then compared to human expert labels.

Two sets of performance metrics are calculated, metrics based on the success/failure in classifying individual 16~sec long segments, and event-based metrics which indicate whether a seizure is detected at all, or whether a seizure is falsely reported. The segment-based metrics are sensitivity (SE), specificity (SP) and the area under the receiver operating characteristic curve (AUC). These metrics are calculated from segments without disagreements between human experts and segments with either seizure either non-seizure activity for the whole segment duration. The event-based metrics are seizure detection rate (SDR), false detections per hour (FD/h) and the mean false detection duration (MFDD)~\cite{temko2011performance}. A consensus seizure is considered to be detected if it is detected at any point in time and a seizure is considered as a false detection if it did not overlap with any (consensus or not) seizure labelled by the human experts. Definitions of the metrics are provided in appendix~\ref{app: performanceMetrics}. Metrics calculated on each patient separately are summarized by their means and medians.

Before the event-based metrics are calculated a post-processing step is in order since segments overlap. Besides a few segments at the beginning and end of each recording, for each 4~sec long segment there are 4 overlapping 16~sec long segments. Prediction for a 4~sec segment is obtained by averaging predictions from overlapping 16~sec long segments~\cite{ingolfsson2021towards, pale2021systematic}. Seizures with duration less than 10~sec are excluded and considered normal brain activity as by definition seizures are longer than 10~sec~\cite{tsuchida2013american}.

For studying the segment-based level of agreement between the local NSDAs we use Gwet's first-order agreement coefficient (AC1)~\cite{gwet2014handbook}. Compared to the often used Cohen's (Fleiss')~$\kappa$~\cite{isaev2020attention, borovacinfluence, stevenson2019hybrid}, Gwet's AC1 is less prone to the paradoxes associated with highly imbalanced data~\cite{feinstein1990high, wongpakaran2013comparison}. 

Performance on the 3-channel DS is evaluated in the same manner as for the 18-channel DS, i.e.\ the metrics are calculated for each patient separately and then summarized with the mean and the median. The baseline NSDA is trained using all 38 patients in $P$ (no patients are left-out), and the union of the training sets for the local NSDAs also contain all 38 patients in $P$. This result in additional $1 + 10 \cdot (3 + 4 + \cdots + 10) = 521$ NSDAs and $10 \cdot (1 +1 + \cdots + 1) = 10 \cdot 8 = 80$ logistic regression classifiers.

\section{Results}

To assess the clinical usefulness of the aggregation schemes they are compared to a baseline NSDA which is trained on data from all 38 patients in $P$ (in a leave-one-subject-out setting for evaluation on the 18-channel DS). The baseline NSDA thus corresponds to the situation where a single agent has access to all the training data ($P$), a situation which is expected to be favorable compared to aggregating predictions from multiple models trained on disjoint subsets of the same data. 

\subsection{Baseline NSDA}
\label{subsec: baseModel}

Table~\ref{tab: baseLiterature} compares the performance of the baseline detector to other NSDAs found in the literature. All detectors are neural networks and were trained or tested using the 18-channel DS. The difference between the mean (0.92) and median (0.98) AUC values for the baseline NSDA calculated on the 18-channel DS is mainly due to the presence of respiratory and heart rate artefacts and low seizure burden in some of the recordings.

\begin{table}[h!]
    \centering
    \caption{Comparison of the area under the curve (AUC) values found in the literature. Each reference uses a different proprietary data set. All NSDAs, except~\cite{isaev2020attention}, were trained using the 18-channel DS. Superscript L denotes leave-one-subject-out testing and superscript C denotes AUC value on concatenated recordings from the data set.}
    \label{tab: baseLiterature}
    \begin{tabular}{ll S[table-format = 1.2] S[table-format = 1.2]}
        \toprule
        && \multicolumn{2}{c}{AUC} \\
        \cmidrule{3-4}
        && \multicolumn{1}{c}{18-channel DS} & \multicolumn{1}{c}{Proprietary DS} \\ \midrule
        Isaev et al.~\cite{isaev2020attention} & mean & 0.92 & 0.97$^{\text{L}}$ \\
        O'Shea et al.~\cite{o2020neonatal} & mean & 0.96$^{\text{C}}$ & 0.99$^{\text{L}}$ \\  
        Stevenson et al.~\cite{stevenson2019hybrid} & median & 0.99$^{\text{L}}$ \\ \midrule
        \multirow{2}{*}{Baseline NSDA} & median & 0.98$^{\text{L}}$ & 0.93 \\
        & mean & 0.92$^{\text{L}}$ & 0.92 \\
        \bottomrule
    \end{tabular}
\end{table}
The performance of an NSDA on an independent test set is usually worse than performance estimates obtained from a held out training data. Such a decrease can be attributed to several factors, including differences in patient cohorts, seizure prevalence, the number of available EEG channels, the human experts that annotated the EEG~\cite{borovacinfluence}, and training data not representing the general population. For example, the mean AUC decreased from 0.97 to 0.92 in~\cite{isaev2020attention} and from 0.99 to 0.96 in~\cite{o2020neonatal}. We observe a similar drop in performance when the baseline detector was tested on a proprietary the 3-channel DS. Detailed validation of the NSDA performance is available in table~\ref{tab: baseClassifierEvents} in appendix~\ref{app: additionalResults}.

In summary, the baseline NSDA gives comparable results to the state-of-the-art NSDAs and performs well on recordings which include only a small subset of the channels used in training. 

\subsection{Aggregation schemes}
\label{subsec: resultsAggregationMethods}

Here we evaluate the different aggregation schemes and compare them to the baseline NSDA and to the average performance of the local NSDAs. If the baseline performance can be reached with an aggregation scheme, it would indicate that the data does not need to be shared during the training of an NSDA to obtain a detector with state-of-the-art performance. 
The four aggregation schemes, majority vote, mean, weighted mean and the Dawid--Skene method were evaluated on the 18-channel DS and the 3-channel DS for $k=3,4,\ldots, 10$ local NSDAs. Results for the majority vote are not shown since in all cases majority vote was slightly outperformed by the mean aggregation scheme (see figure~\ref{fig: randomMV} in appendix~\ref{app: additionalResults}).

\begin{figure*}[h!]
    \centering
    \includegraphics[width = \textwidth]{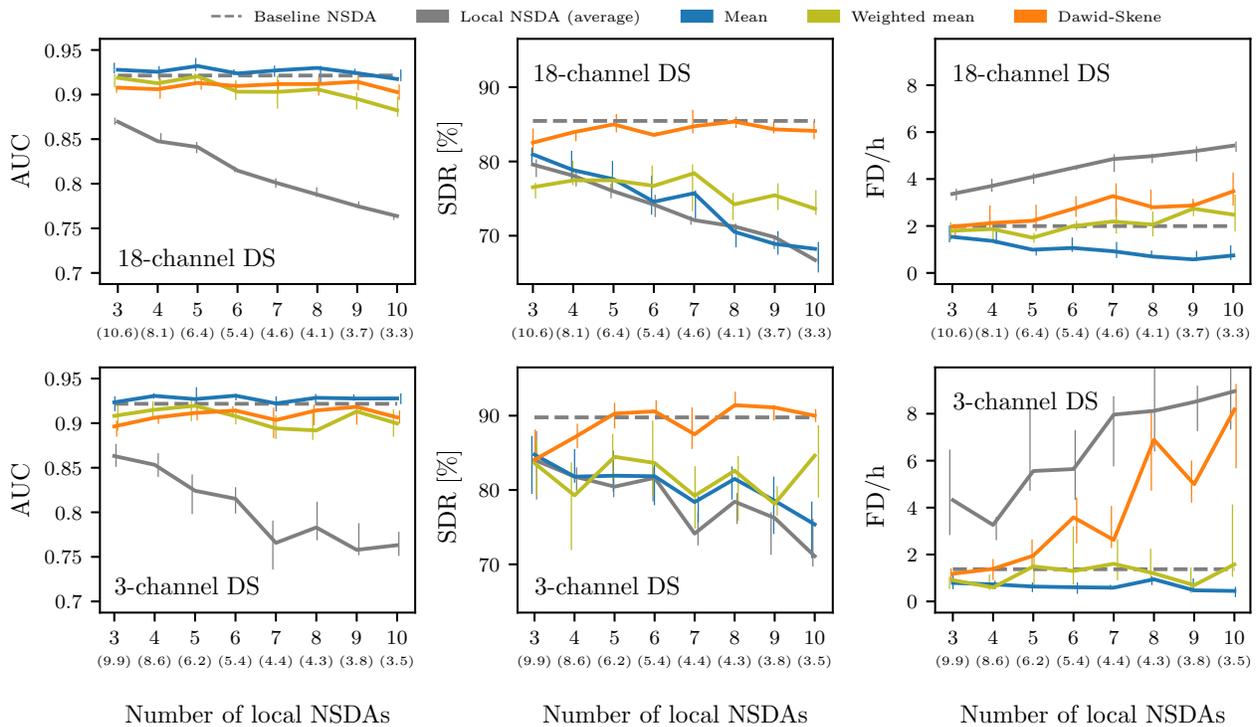}
    \caption{Average area under the curve (AUC), seizure detection rate (SDR) and false detections per hour (FD/h) as a function of the number of local NSDAs used in the aggregation schemes. The solid lines represent the medians of ten runs together with interquartile ranges denoted with vertical lines. The grey dashed line represents the average metric of the baseline NSDA. The average (across ten runs) mean median number of patients in each NSDA is shown in parentheses.}
    \label{fig: randomSDRFDMFDD}
\end{figure*}

With an increasing number of local NSDAs the average performance of an individual detector gradually gets worse (figure~\ref{fig: randomSDRFDMFDD}). This is explained by the fact that the number of patients behind each local NSDA is becoming smaller since the total number of patients in the combined training sets is constant (37 for the 18-channel DS and 38 for the 3-channel DS). Consequently there is an increased risk of overfitting in individual detectors. The size of the local training sets is quantified with the mean median number of patients in the training set. E.g.,\ if four local NSDAs are used and the mean median is 8.1, then on average there are at least nine patients in the training of two of the local NSDAs.

Figure~\ref{fig: randomSDRFDMFDD} shows that the AUC, seizure detection rate and false detection rate behave similarly across both data sets for all the aggregation schemes, but there is considerably more variability for the 3-channel DS.
All the aggregation schemes give AUC values that are similar to the baseline value. However, the aggregation schemes differ in terms of seizure detection rate and false detections per hour.

\begin{figure}[h]
    \centering
    \includegraphics{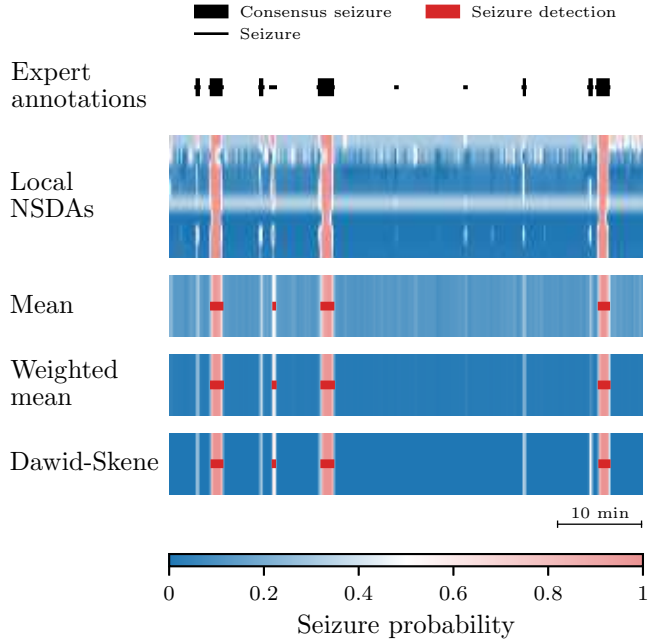}
    \caption{An example of aggregated predictions from eight local NSDAs. The area under the curve is 1.0 for the mean and the weighted mean and 0.99 for the Dawid--Skene method. All aggregation schemes detect 42.9~\% of consensus seizures and they do not falsely detect any seizure.}
    \label{fig: exampleHighFDDSAggregation}
\end{figure}

Figure~\ref{fig: exampleHighFDDSAggregation} shows the seizure probability estimates returned by local NSDAs for an hour-long recording, together with probability estimates obtained with the ensemble methods. All the aggregation schemes result in AUC close to one, although they detect only 3 out of 7 consensus seizures. The missed seizures are short in duration and they are clearly visible in the figure (as white bands) since the corresponding probabilities are higher than for the non-seizure segments. 

The SDR in figure~\ref{fig: randomSDRFDMFDD} behaves similarly for both data sets. For all values of $k$ tested, the Dawid--Skene method is comparable to the baseline NSDA, while for the mean and the weighted mean aggregation schemes, fewer seizures were detected with an increased number of local NSDAs. Recall that when there are few NSDAs, each NSDA detects almost as many seizures as the baseline detector. The mean aggregation scheme performed slightly worse than the weighted mean and both performed notably worse than the Dawid--Skene method for more than four local detectors. Moreover, the average SDR of the local NSDAs is comparable to the values corresponding to the mean aggregation scheme. With the weighted mean a larger number of seizures are detected for $k \geq 8$ ($k = 10$) on the 18-channel DS (3-channel DS), for smaller $k$ the mean and the weighted mean aggregation schemes return comparable seizure detection rates.

Moreover, in figure~\ref{fig: randomSDRFDMFDD} we observe that all aggregation schemes result in a lower number of FD/h than the average local NSDA. The average FD/h of the local NSDAs are noticeably higher for the 3-channel DS than for the 18-channel DS. One possible explanation is that the recordings in the 3-channel DS are much longer and on average just 3.5~\% of a recording corresponds to a seizure activity. The mean aggregation scheme has a lower false detection rate than the baseline NSDA and the FD/h decreases steadily with increasing number of local NSDAs. This may be a result of low level of agreement between the local NSDAs for the large $k$ (figure~\ref{fig: gwet} in appendix~\ref{app: additionalResults}). So, even though an individual local NSDA falsely detects a large number of seizures, the aggregated prediction filtered them out or was below the 0.5 threshold. This may on the other hand caused problems with the Dawid--Skene method, i.e.\ the FD/h increased slowly on the 18-channel DS and rapidly on the 3-channel DS with increasing number of local NSDAs. In contrast, the logistic regression classifier determining the weights for the weighted mean aggregation scheme successfully detected local NSDAs with high/low false detection rate for all $k$ tested.

We observed low false detection rates for the mean and weighted mean aggregation schemes and therefore investigated whether the false detections are short or long in duration. We did not observe big differences between the aggregation schemes (10 - 30~sec) and different values of local NSDAs (figure~\ref{fig: randomMFDD} in appendix~\ref{app: additionalResults}).

To summarise, all aggregation schemes tested here performed better than the average local NSDA and were comparable to the baseline NSDA for $k \in \{3, 4\}$. This shows that the overfitting by local models noted earlier is offset by aggregating their predictions. This is in line with published reports on ensemble methods such as Random Forests which aggregate predictions from multiple models individually overfitting the data. The decrease in performance for larger values of $k$ is mainly a result of training the local NSDAs on smaller training sets that do not capture the general population. The (weighted) mean aggregation scheme detects fewer seizures than the baseline detector, however the false detection rate is comparable, if not lower. The Dawid--Skene method successfully detects the same number of seizures as the baseline NSDA for any number of local NSDAs, but the false detection rate is compromised for $k \geq 6$. Predictions obtained with the Dawid--Skene are difficult to  explain~\cite{ibrahim2019crowdsourcing, zhang2016spectral}, only a few local NSDAs with poor performance may have caused unexpected and undesired aggregated prediction~\cite{miao2018attack}.

\section{Conclusion}

In this work we have shown that an NSDA based on a convolutional neural network together with an attention layer can accurately detect seizures, even if the data is obtained with different types of electrodes (scalp vs needle) and significantly lower number of channels than it was used for training. 
All the performance metrics of the NSDAs unsurprisingly dropped when training sets contained data from only a few patients. For aggregation of such NSDAs the weighted mean aggregation scheme performed best. Compared to the Dawid--Skene method, it successfully detected local NSDAs with high false detection rates and seizure detection rate was not as compromised as it was for the mean aggregation scheme.
When a larger number of patients was included in the training of individual local NSDAs, i.e.\ when the number of local NSDAs was few, the Dawid--Skene method marginally outperformed the other aggregation schemes. It had a higher seizure detection rate and the false detections per hour was comparable to the (weighted) mean aggregation scheme.
Independent of the number of local NSDAs, the majority vote was slightly outperformed by the mean aggregation scheme and all aggregation schemes performed better than the average individual (local) NSDA. 

The experiments suggest that data does not need to be shared between institutions. It takes approx.\ 15 seconds to process one hour of 18-channel EEG with 10 local detectors, which is fast enough to be used in an online setting in the clinic. By utilizing GPU optimized code in the preprocessing steps and a fast version of the Dawid-Skene aggregation method~\cite{sinha2018fast}, one hour of EEG could be processed in less than 2 seconds.

To confirm the findings reported here in a real-world setting, data from multiple institutions would be required. A large data set would also allow a detailed study on the number of local NSDAs needed to reach the desirable classification performance and whether a mixture of different types of NSDAs improves or degrades the overall performance.

\section*{Acknowledgment}
This project receives funding from the Sigrid Juselius Foundation and the European Union’s Horizon 2020 research and innovation programme under grant agreement No 813483.

\bibliographystyle{plain}
\bibliography{references}

\begin{appendices}

\section{Dawid--Skene method}
\label{app: dawidSkene}

The Dawid--Skene method was initially used to estimate the performance of human annotators~\cite{dawid1979maximum}. Here the method is used to estimate the performance of models (local NSDAs) and obtain consensus judgement amongst them. The method is as follows.
From a given set $D$ of model predictions, the task is to estimate consensus labels $\{ \mu_i \}_{i = 1}^N$, the sensitivity $\alpha^j$ and specificity $\beta^j$ of predictive model $j \in \{1, 2, \dots, R\}$.
Let $Y_e$ denote the multivariate random variable
\[
Y_e = (Y_1^1, Y_1^2, \dots, Y_1^R, \dots, Y_N^1, Y_N^2, \dots, Y_N^R),
\]
where random variable $Y_i^j$ denotes the label given to instance $i$ by model $j$. Furthermore, let $T_i$ denote a random variable corresponding to the true label of instance $i$ for which 
\[
P[T_i = 1] = t_i = t;\quad i \in \{ 1, 2, \dots, N \}.
\]
Assuming that model labels are independent and that conditional probability of $Y_i^j$ on $T_i$ follows Bernoulli distribution with parameters $\alpha^j$ and $\beta^j$, respectively:
\begin{align*}
    a_i & = P_\alpha \left[ Y_i^1, Y_i^2, \dots, Y_i^R | T_i = 1 \right] \\
    & = \prod_{j = 1}^R (\alpha^j)^{y_i^j} (1 - \alpha^j)^{1-y_i^j};\quad i \in \{ 1, 2, \dots, N \}, \\
    b_i & = P_\beta \left[ Y_i^1, Y_i^2, \dots, Y_i^R | T_i = 0 \right] \\
    & = \prod_{j = 1}^R (\beta^j)^{1 - y_i^j} (1 - \beta^j)^{y_i^j};\quad i \in \{ 1, 2, \dots, N \}.
\end{align*}

To simplify the notation, let $\theta = (t, \alpha, \beta)$ denote the parameters to be estimated. Assuming that instances are sampled independently, the likelihood function for $Y_e$ is~\cite{dawid1979maximum, raykar2010learning}:

\begin{align}
P_\theta[Y_e] & = \prod_{i = 1}^N P_\theta[ Y_i^1, Y_i^2, \dots, Y_i^R] \notag \\
 & = \prod_{i = 1}^N \left( \underbrace{P_\theta[ Y_i^1, Y_i^2, \dots, Y_i^R | T_i = 1]}_{a_i} \underbrace{P_\theta[T_i = 1]}_{t} \right. \notag \\
 & \left. \hspace{0.92cm} + \underbrace{P_\theta[Y_i^1, Y_i^2, \dots, Y_i^R | T_i = 0]}_{b_i} \underbrace{P_\theta[T_i = 0]}_{1 - t} \right) \notag \\
& = \prod_{i = 1}^N \left( a_it + b_i(1-t) \right) . \label{eq: likelihood} 
\end{align}

Dawid and Skene used the EM algorithm to identify a local maximum of the likelihood function. The true labels are estimated by maximizing the likelihood function using estimated values for the sensitivity and specificity of each annotator, and the prior probability of class 1 ($t$), i.e.\ seizure. The algorithm has two main steps~\cite{dawid1979maximum}.

Expectation step: calculate the expected value of a true label knowing labels made by predictive models
    \begin{align}
        \mu_i & = \mathbb{E} [T_i | Y_i^1, Y_i^2, \dots, Y_i^R] \notag \\ 
        & = P_\theta[T_i = 1 | Y_i^1, Y_i^2, \dots, Y_i^R] \notag \\
        & = \frac{P_\theta[Y_i^1, Y_i^2, \dots, Y_i^R | T_i = 1] P_\theta[T_i = 1]}{P_\theta[Y_i^1, Y_i^2, \dots, Y_i^R]} \tag{Bayes' theorem} \\
        & = \frac{a_it}{a_it + b_i (1 - t)};\quad i \in \{1, 2, \dots, N\}. \label{eq: mu}
    \end{align}

Maximization step: estimate $t$, $\alpha^j$ and $\beta^j$ that maximize the likelihood function~\eqref{eq: likelihood}
    \begin{align}
        t & = \frac{\sum_{i = 1}^N \mu_i}{N}, \label{eq: p} \\
        \alpha^j & = \frac{\sum_{i = 1}^N \mu_i y_i^j}{\sum_{i = 1}^N \mu_i}; \quad j \in \{1, 2, \dots, R\}, \label{eq: alpha} \\
        \beta^j & = \frac{\sum_{i = 1}^N (1 - \mu_i) (1 - y_i^j)}{\sum_{i = 1}^N ( 1 - \mu_i)}; \quad j \in \{1, 2, \dots, R\}. \label{eq: beta}
    \end{align}
In the special case when all the $\mu_i$'s are either 0 or 1, then $t$ is the estimated ratio of positive instances and $\alpha^j$ ($\beta^j$) is an estimated ratio of correctly predicted positive (negative) examples by expert $j$, i.e.\ the estimated sensitivity (specificity) of expert $j$.

\vspace{1.5 mm}
\begin{algorithmic}
\REQUIRE $D$, $\epsilon = 10^{-5}$, $k_{max} = 5000$
\ENSURE $\mu^{D\!S}$
\STATE initialize $\mu^{D\!S} = \mu^M$
\STATE compute $\theta^{(0)}$ using equations~\eqref{eq: p}, \eqref{eq: alpha} and \eqref{eq: beta}
\STATE $k = 0$
\REPEAT 
\STATE k = k + 1
\STATE compute $\mu^{D\!S}$ using equation~\eqref{eq: mu}
\STATE compute $\theta^{(k)}$ using equations~\eqref{eq: p}, \eqref{eq: alpha} and \eqref{eq: beta}
\UNTIL $|\log P_{\theta^{(k-1)}}[Y_e] - \log P_{\theta^{(k)}}[Y_e]| < \epsilon \quad {\bf or} \quad k \geq k_{max}$
\end{algorithmic}
\vspace{1.5 mm}

\newpage
\section{Data information}
\label{app: data}

\begin{table}[h]
\renewcommand{\arraystretch}{1.5}
    \centering
    \caption{A summary of the data sets used in the study. Numbers inside parentheses represent standard deviation. Means for recordings are calculated across patients containing at least one consensus seizure longer than 16~sec (duration of one EEG segment).}
    \label{tab: basicInfo}
    \begin{tabular}{lcc}
    \toprule
        & 18-channel DS & 3-channel DS  \\ \midrule \midrule
        Number of patients & $79$ & $28$ \\
        Number of patients with consensus seizures $\geq 16~$sec  & $38$ & $24$ \\
        Gestational age (weeks) & $39.3$ ($2.1$) & $39.2$ ($2.0$) \\
        Number of derived EEG channels & $18$ & $3$ \\
        Total recordings duration (hours) & $111.9$ & $2149.4$ \\
        Mean recording duration (hours) & $1.4$ ($0.6$) & $76.4$ ($35.8$) \\
        Total number of consensus seizures & $344$ & $1387$ \\
        Total duration of consensus seizures (hours) & $11.0$ & $65.3$ \\
        Mean duration of consensus seizures (minutes) & $1.9$ ($2.7$) & $2.8$ ($6.0$) \\
        Mean fraction of recording containing seizures (\%) & $31.8$ ($26.4$) & $5.3$ ($5.6$) \\
        Mean fraction of recording containing consensus seizures (\%) & $19.1$ ($20.9$) & $3.5$ ($4.0$) \\
    \bottomrule
    \end{tabular}
\end{table}

\newpage
\section{Architecture of the NSDA}
\label{app: dnn}

\begin{figure}[h]
    \centering
    \includegraphics[width = 0.6\columnwidth]{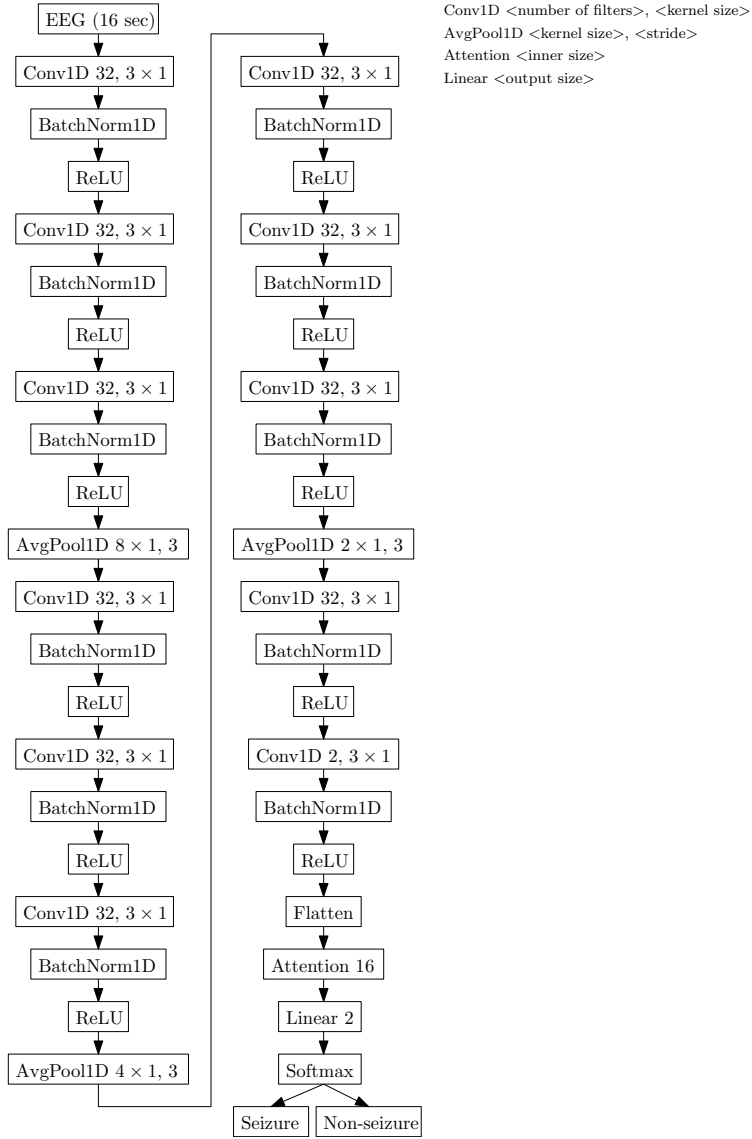}
    \caption{Architecture of the NSDA with a total of 29352 learnable parameters. Other parameters were set to default PyTorch values.}
    \label{fig: nn}
\end{figure}

In this work the NSDAs are deep neural networks consisted of three components, a feature extractor~\cite{o2018investigating}, an attention layer~\cite{isaev2020attention} and an output layer (figure~\ref{fig: nn}). We used PyTorch implementation of layers for the feature extractor and for the output layer. Using PyTorch notation, the attention layer was implemented as follows. If an input to the attention layer is of size $(N, C_{in}, L)$ then the output is of size $(N, L)$ and can be described as
\begin{align*}
    & \text{out}(N_i) = \sum_{k = 0}^{C_{in}-1} a_k \text{input}(N_i, k); \\
    & a_k = \frac{\exp \left( w^T \tanh \left( V \text{input}(N_i, k)^T \right) \right) }{\sum_{j = 0}^{C_{in}-1} \exp \left( w^T \tanh \left( V \text{input} (N_i, j)^T \right) \right) },
\end{align*}
where $V \in \mathbb{R}^{L \times <\text{inner size}>}$ and $w \in \mathbb{R}^{L \times 1}$ are learnable parameters.

\newpage
\section{Performance metrics}
\label{app: performanceMetrics}

\subsection*{Segment-based metrics}

Segment-based metrics were calculated based on 16~sec long EEG segments. A true positive (TP) is a correctly predicted seizure segment, a true negative (TN) is a correctly predicted non-seizure segment, a false positive (FP) is an incorrectly predicted non-seizure segment and a false negative (FN) is an  incorrectly predicted seizure segment.
\begin{itemize}
    \item Sensitivity (ratio of correctly predicted seizure intervals):
    \[
    \SE = \frac{\TP}{\TP + \FN} \cdot 100.
    \]
    \item Specificity (ratio of correctly predicted non-seizure intervals):
    \[
    \SP = \frac{\TN}{\TN + \FP} \cdot 100.
    \]
    \item Area under the receiver operating characteristics curve (AUC). The receiver operating characteristics curve describes SE depending on 1-SP.
\end{itemize}

\subsection*{Event-based metrics}

Event-based metrics are in comparison with the segment-based metrics focused on each predicted seizure and not just 16~sec long segments. Three event-based metrics were used~\cite{temko2011performance}:
\begin{itemize}
    \item Seizure detection rate (SDR):
    \[
    \text{SDR} = \frac{\text{DS}}{\text{CS}} \cdot 100,
    \]
    where DS is a number of detected consensus seizures and CS is a number of consensus seizures. A seizure was considered to be detected if it was detected at any time of its duration. 
    \item False detections per hour (FD/h):
    \[
    \text{FD/h} = \frac{\text{IDS}}{\text{D}},
    \]
    where IDS is a number of incorrectly detected seizures and D is duration of data in hours. A seizure was considered to be incorrectly detected if it is not overlapping with any seizure annotated by the experts.
    \item Mean false detection duration (MFDD):
    \[
    \text{MFDD} = \begin{cases} 0; & \text{if IDS}=0  \\
    \frac{\text{DIDS}}{\text{IDS}}; & \text{otherwise}
    \end{cases},
    \]
    where DIDS is a sum of durations of incorrectly detected seizures in seconds and IDS is a number of incorrectly detected seizures.
\end{itemize}

\newpage
\section{Additional results}
\label{app: additionalResults}

\begin{table}[h]
    \centering
    \caption{Accuracy of the baseline model. Area under the curve (AUC), sensitivity (SE), specificity (SP), seizure detection rate (SDR), false detections per hour (FD/h) and mean false detection duration (MFDD) are computed as the mean and median over all the patients with seizures.}
    \label{tab: baseClassifierEvents}
    \begin{tabular}{ll ccc}
        \toprule
        && \multicolumn{3}{c}{Segment-based metrics} \\ \cmidrule{3-5}
        & & \multicolumn{1}{c}{AUC} & \multicolumn{1}{c}{SE [\%]} & \multicolumn{1}{c}{SP [\%]} \\ \midrule 
        \midrule
        \multirow{2}{*}{18-channel DS} & median & $0.98$ & $90.46$ & $97.21$ \\
        & mean & $0.92$ & $79.52$ & $93.69$ \\ \midrule 
        \multirow{2}{*}{3-channel DS} & median & $0.93$ & $78.00$ & $98.23$ \\
        & mean & $0.92$ & $70.54$ & $97.40$ \\ \bottomrule \toprule
        && \multicolumn{3}{c}{Event-based metrics} \\ \cmidrule{3-5}
        && \multicolumn{1}{c}{SDR [\%]} & \multicolumn{1}{c}{FD/h} & \multicolumn{1}{c}{MFDD [s]} \\ \midrule \midrule
        \multirow{2}{*}{18-channel DS} & median & $100.0$ & $0.91$ & $12.00$ \\
        & mean & $85.45$ & $1.99$ & $19.15$ \\ \midrule 
        \multirow{2}{*}{3-channel DS} & median & $95.55$ & $0.97$ & $15.82$ \\
        & mean & $89.77$ & $1.37$ & $17.92$ \\
        \bottomrule
    \end{tabular}
\end{table}

\begin{figure}[h]
    \centering
    \includegraphics{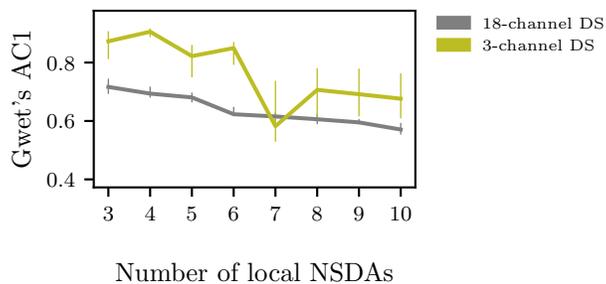}
    \caption{Average Gwet's AC1 between local NSDAs for 18-channel DS and 3-channel DS. The solid lines represent the medians of ten runs together with interquartile ranges denoted with vertical lines.}
    \label{fig: gwet}
\end{figure}

\begin{figure*}[h]
    \centering
    \includegraphics[width = \textwidth]{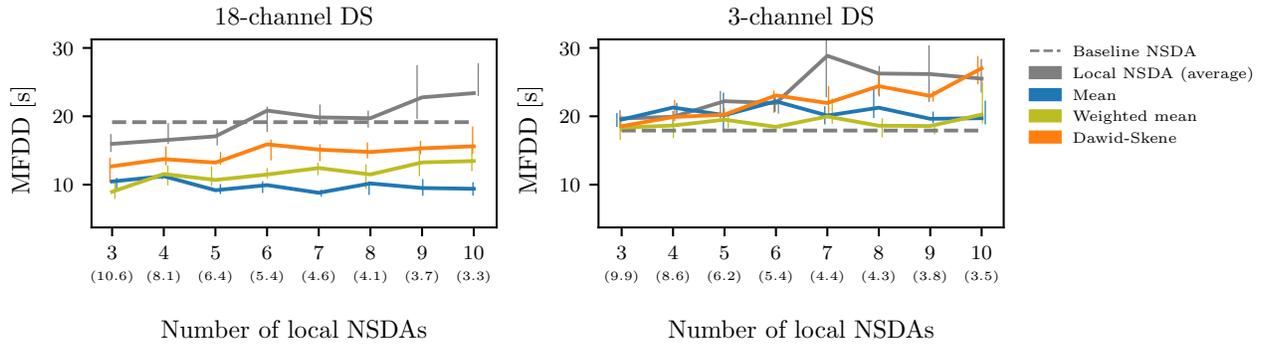}
    \caption{Average mean false detection duration (MFDD) as a function of the number of local NSDAs used in the aggregation schemes. The solid lines represent the medians of ten runs together with interquartile ranges denoted with vertical lines. The grey dashed line represents the average MFDD of the baseline NSDA.}
    \label{fig: randomMFDD}
\end{figure*}

\begin{figure*}[h]
    \centering
    \includegraphics[width = \textwidth]{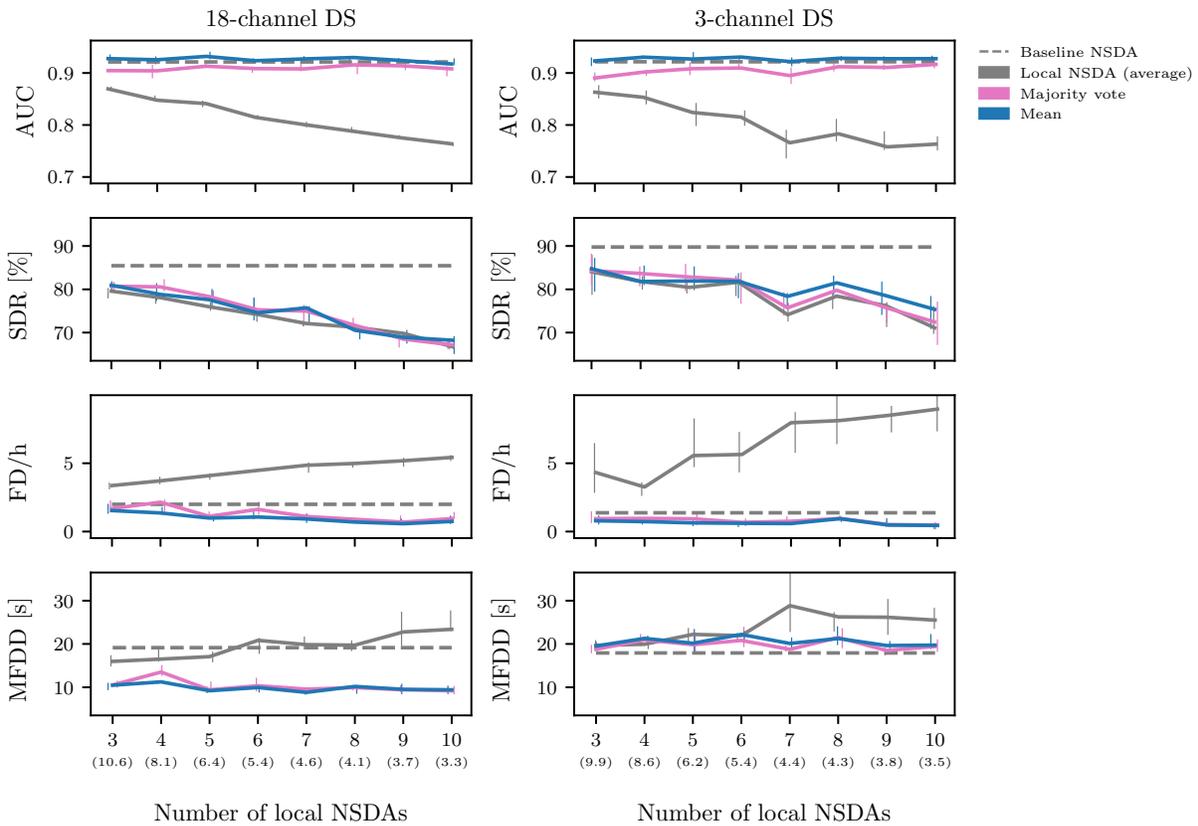}
    \caption{Average area under the curve (AUC), seizure detection rate (SDR), false detections per hour (FD/h) and false detection duration (MFDD) as a function of the number of local NSDAs used in the aggregation schemes. The solid lines represent the medians of ten runs together with interquartile ranges denoted with vertical lines. The grey dashed line represents the average metric of the baseline NSDA. The average (across ten runs) mean median number of patients in each NSDA is shown in parentheses.}
    \label{fig: randomMV}
\end{figure*}

\end{appendices}
\end{document}